\documentclass[aps,showpacs,10pt,pra,amsmath,amssymb,superscriptaddress,twocolumn]{revtex4-1}
\usepackage{graphicx}
\usepackage{hyperref}
\usepackage{dcolumn}
\usepackage{bm}
\usepackage{color}
\usepackage{graphicx}
\usepackage{ulem}
\usepackage{natbib}

\begin{document}

\title{Invariant currents in lossy acoustic waveguides with complete local symmetry}

\author{P.A. Kalozoumis}
\affiliation{Zentrum f\"ur Optische Quantentechnologien, Universit\"{a}t Hamburg, Luruper Chaussee 149, 22761 Hamburg, Germany}
\affiliation{Department of Physics, University of Athens, GR-15771 Athens, Greece}

\author{O. Richoux}
\affiliation{LUNAM Universit\'e, Universit\'e du Maine, CNRS, LAUM UMR 6613, Avenue O. Messiaen,
72085 Le Mans, France}

\author{F.K. Diakonos}
\affiliation{Department of Physics, University of Athens, GR-15771 Athens, Greece}

\author{G. Theocharis}
\affiliation{LUNAM Universit\'e, Universit\'e du Maine, CNRS, LAUM UMR 6613, Avenue O. Messiaen,
72085 Le Mans, France}

\author{P. Schmelcher}
\affiliation{Zentrum f\"ur Optische Quantentechnologien, Universit\"{a}t Hamburg, Luruper Chaussee 149, 22761 Hamburg, Germany}
\affiliation{The Hamburg Centre for Ultrafast Imaging, Universit\"{a}t Hamburg, Luruper Chaussee 149, 22761 Hamburg, Germany}

\date{\today}

\begin{abstract}

We implement the concept of complete local symmetry in lossy acoustic waveguides.
Despite the presence of losses, the existence of a spatially invariant current is shown theoretically and observed experimentally. We demonstrate how this invariant current leads to the generalization of the Bloch and parity theorems for lossy systems defining a mapping of the pressure field between symmetry related spatial domains.  Using experimental data we verify this mapping with remarkable accuracy. For the performed experiment we employ a construction technique based on local symmetries which allows the design of setups with prescribed perfect transmission resonances in the lossless case. Our results reveal the fundamental role of symmetries in restricted spatial domains and clearly indicate that completely locally symmetric devices constitute a promising class of setups, regarding the manipulation of wave propagation. 
\end{abstract}

\maketitle

\section{Introduction}\label{intro}

Symmetries play a fundamental role in our search for
the physical laws in nature. Significant theoretical and phenomenological advances have been obtained using models  characterized by global symmetries, including the existence of forbidden bands of wave propagation in infinite periodic structures (Bloch theorem)~\cite{Bloch1929}
and the existence of even or odd eigenstates in problems that involve potential with global inversion symmetry (parity theorem)~\cite{Zettili2009}. However, global symmetry and its consequences constitutes an idealized scenario, requiring often a sophisticated preparation procedure to be experimentally probed.  Opposite to this, a frequently encountered scenario in nature concerns complex systems  with a variety of symmetries valid in limited spatial domains of the entire embedding space~\cite{Echeverria2011}. For their characterization we apply here the term \textit{local symmetries} and
their emergence can be attributed to mechanisms of global symmetry breaking leading to new symmetries (remnants) at different scales.

Indeed, such spatially localized symmetries can be intrinsic in complex physical systems like large molecules~\cite{Grzeskowiak1993,Pascal2001},
quasicrystals~\cite{Shechtman1984,Widom1989,Verberck2014}, or even in partially disordered matter~\cite{Wochner2009}.
Furthermore, since technological advances often require tailored structures suitable for corresponding applications,
local symmetries can be present by design in multilayered photonic devices~\cite{Macia2006,Zhukovsky2010,Peng2002},
semiconductor superlattices~\cite{Ferry1997} and magnonic systems~\cite{Hsueh2013}.  Acoustic~\cite{Hao2010,Aynaou2005,King2007} and phononic~\cite{Hladky2013,Tamura1987,Mishra2004} structures have attracted increasing interest in this direction. 
Despite the fact that systems belonging to the aforementioned classes have been extensively investigated 
\cite{Kohler2002,Wu2004,Scotognella2015,Fung2009,Zhu2011}, the impact of their local symmetries has not been explored thoroughly.

The natural step beyond the above ascertainment would be to investigate whether -and how- these  locally existing symmetries carry imprints of the equivalent global symmetries and in turn, by which procedure these could be used to derive structural characteristics of the scattered wave field.  
Recently, we developed a rigorous formalism~\cite{Kalozoumis2014a} in order to describe the impact of local symmetries in  one-dimensional, lossless scattering  setups.
To this aim we employed a generic wave mechanical framework -equally applicable to acoustics, photonics and quantum mechanics- described by the generalized Helmholtz equation $\mathcal{A}''(x)+U(x)\mathcal{A}(x)=0$, where $\mathcal{A}(x)$ is the complex wave field and $U(x)$ is the potential which describes the inhomogeneities of the medium. For the case of acoustic waves, on which this work focuses, $\mathcal{A}(x)$ corresponds to the pressure field $p(x)$.

\begin{figure}[hbt]
\centering
\includegraphics[width=0.99\columnwidth]{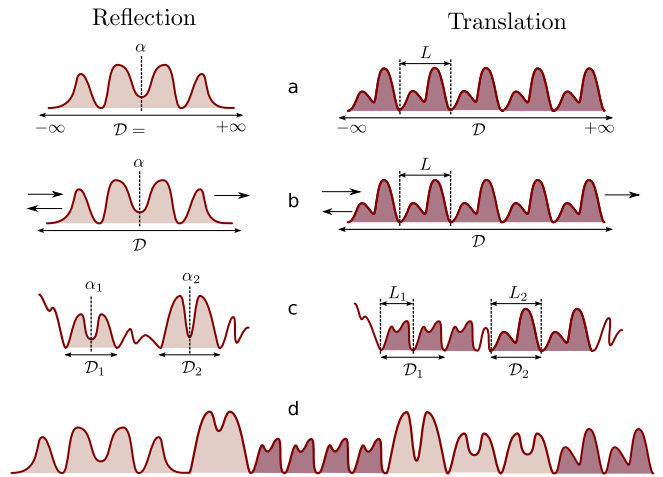}
\caption{ (Color online) (a) Globally symmetric system. The corresponding symmetry is obeyed $\forall ~x \in ~\mathbb{R}$ and the wave field is an eigenfunction of the symmetry operator. (b) Even though the potential itself is symmetric $\forall ~x \in ~\mathcal{D}$, the asymmetric asymptotic conditions break the global symmetry of the system. (c) The potential is not symmetric $\forall ~x \in ~\mathcal{D}$. Nevertheless, symmetry is retained locally in the domains $\mathcal{D}_{1},~\mathcal{D}_{2}$. (d)  Completely locally symmetric (CLS) setup, consisting exclusively of potential domains with local symmetries. 
}
\label{fig_new}
\end{figure}

In one dimension, the basic symmetry transformations are reflection through a point $\alpha$ and translation by a length $L$, both described by the form $F(x) \equiv \bar{x}=\sigma x + \rho$, allowing for a unified description as: (i) Reflection: $\sigma=-1,~\rho=2\alpha$, (ii) Translation:  $\sigma=1,~\rho=L$.
 Let us assume that $U(x)$ obeys the symmetry $U(x)=U(F(x))$ within an arbitrary domain $\mathcal{D} \subseteq \mathbb{R}$. If $\mathcal{D}=\mathbb{R}$ then the symmetry is \textit{global}. In any other case the symmetry will be regarded as \textit{local} and can be of different kinds, as shown in Fig.~\ref{fig_new} (for a detailed description of the different types of local symmetries see~\cite{Kalozoumis2014a}). The symmetry of the potential  leads to a pair of invariant,  symmetry-induced nonlocal currents,
\begin{equation}
\label{Q_prl} Q=\frac{1}{2i} [ \sigma \mathcal{A}(x) \mathcal{A}'(\bar{x}) - \mathcal{A}(\bar{x}) \mathcal{A}'(x) ],
\end{equation}
\begin{equation}
\label{Qtilde_prl} \widetilde{Q}=\frac{1}{2i} [ \sigma \mathcal{A}^{*}(x) \mathcal{A}'(\bar{x}) - \mathcal{A}(\bar{x}) \mathcal{A}'^{*}(x) ],
\end{equation} 
which remain constant within $\mathcal{D}$. 
The term ``nonlocal'' stems from the fact that Eqs.~(\ref{Q_prl}),~(\ref{Qtilde_prl}) involve both symmetry related points $x$ and $\bar{x}$. For a detailed derivation of $Q$ and $\widetilde{Q}$ see Ref.~\cite{Kalozoumis2014a}.

Especially the current $Q$ has a twofold fundamental functionality:  (i) it provides a systematic pathway towards the description of discrete symmetry breaking and (ii) it generalizes~\cite{Kalozoumis2014a} the well-known Bloch~\cite{Bloch1929} and parity~\cite{Zettili2009} theorems for systems with broken translation and reflection symmetry, respectively.

The symmetry of the potential, $U(x)$ $\forall ~x \in \mathbb{R}$, implies the commutation of the Helmholtz operator $\hat{\mathcal{H}}=d^{2}/dx^{2}+U(x)$ with the considered symmetry operator $\hat{\mathcal{O}}_{F}$. Interestingly, by setting $Q=0$ it is straightforward to show that $\mathcal{A}(\bar{x}) = \hat{\mathcal{O}}_{F} \mathcal{A}(x) = c \mathcal{A}(x)$, which implies that the wave field $\mathcal{A}(x)$ becomes an eigenfunction of the respective symmetry operator $\hat{\mathcal{O}}_{F}$. This case corresponds to a globally symmetric system, as shown in Fig.~\ref{fig_new} (a).  On the other hand, Fig.~\ref{fig_new} (b) illustrates a potential which is symmetric ($U(x)=U(\bar{x}),~ \forall~ x \in \mathcal{D}$ where $\mathcal{D}$ is the domain which delimits the potential), though the  symmetry of the system is globally broken due to the asymmetric asymptotic conditions. In this typical scattering situation, even though $[\hat{\mathcal{H}},\hat{\mathcal{O}}_{F}]=0$, the asymptotic conditions prevent the wave field $\mathcal{A}(x)$ from being an eigenfunction of $\hat{\mathcal{O}}_{F}$.  Then while $Q$  is spatially constant within the extent of the setup, its value is different from zero.
Breaking further the symmetry, renders the potential nonsymmetric itself $U(x) \neq U(\bar{x}),~ \forall x \in \mathcal{D}$, and therefore  $[\hat{\mathcal{H}},\hat{\mathcal{O}}_{F}] \neq 0$. If, however, local symmetries are still retained in restricted spatial domains, as  $\mathcal{D}_{1},~\mathcal{D}_{2}$  in Fig.~\ref{fig_new} (c), then $Q_{1},~Q_{2}$ are constant within $\mathcal{D}_{1},~\mathcal{D}_{2}$ and $Q_{1} \neq Q_{2}$, in general. 

Hence, a vanishing $Q$ value characterizes a system with a global, unbroken symmetry, whereas a nonzero $Q$ value signifies the breaking of a symmetry. Note, that in the extreme case of a total symmetry breaking (e.g. strong disorder) $Q=Q(x)$ becomes a function of $x$, lacking the property of constancy, as a result of the lack of any symmetric region residing in the potential landscape. In this sense, a nonvanishing $Q$ value points to the globally broken symmetry, whereas its constancy, evinces the presence of local order and is interpreted as a \textit{remnant} of the corresponding broken global symmetry. This procedure allows to track -in a quantifiable manner- the way from perfect order and global symmetry to aperiodic systems with local symmetries and to disordered systems where no local order is present. 

The power of $Q$ as a symmetry breaking measure, was confirmed in a recent work~\cite{Kalozoumis2014b} where, having the role of the natural order parameter of the system, it allowed for the extension of the phase diagram of the well-known scattering ($\mathcal{PT}$) symmetric dimer~\cite{Chong2011}. Further, studies of scattering systems involving discrete symmetries revealed intriguing properties~\cite{Kalozoumis2013a,Kalozoumis2013b,Morfonios2014} of this pathway to symmetry breaking. 

In the present work we firstly experimentally verify the spatial invariance of the current $Q$ within 
domains of reflection symmetry in aperiodic, completely locally symmetric (CLS)~\cite{Kalozoumis2014a} acoustic waveguides.
The latter can be completely decomposed into domainwise locally symmetric parts (see Fig.~\ref{fig_new} (d)).
We extend the developed theoretical framework in order to incorporate losses and find the surprising fact
that $Q$ remains invariant within the symmetry domains in spite of the presence of losses.
The parity and Bloch theorems are here generalized for inhomogeneous scattering media with losses. As a result a mapping
of the pressure field values between two symmetry related domains is obtained.  Focusing on the case of parity, we
experimentally confirm both the existence of the constant current $Q$ as well as the validity of the associated mapping.
Our aperiodic acoustic waveguide is designed according to the construction principle proposed in Refs.~\cite{Kalozoumis2013a,Kalozoumis2013b}
for CLS setups with prescribed perfect transmission resonances (PTR). The key ingredient allowing for the direct experimental observation 
of the current $Q$ is the ability to measure both the magnitude as well as the phase of the pressure field. 
Therefore CLS acoustic waveguides (and corresponding phononic structures) represent a useful test bed for newly developed wave mechanical concepts.

The paper is organized as follows. In Sec.~\ref{setup-theory} we describe the experimental setup and briefly report on the theory of wave propagation in acoustic waveguides. In Sec.~\ref{setup-design} we describe the procedure followed for the design of the device and the way that losses affect the expected outcome.
Subsequently the existence of the symmetry-induced, spatially invariant current $Q$ is proven in the presence of losses and we provide the corresponding mapping relations which constitute the generalization of the parity and Bloch theorems. Sec.~\ref{results} contains the results of our analysis, including the comparison of the transmission diagrams between the lossless theory and the experiment, the  demonstration of the pressure field spatial profiles, the experimental confirmation
of the invariance of $Q$ and the resulting pressure field mapping between symmetry related
spatial domains. Finally, Sec.~\ref{conclusions} provides our conclusions.

\section{Acoustic waveguide with side-loaded point scatterers}\label{setup-theory}

\subsection{Experimental setup}

\begin{figure*}[hbt]
\centering
\includegraphics[width=2.0\columnwidth]{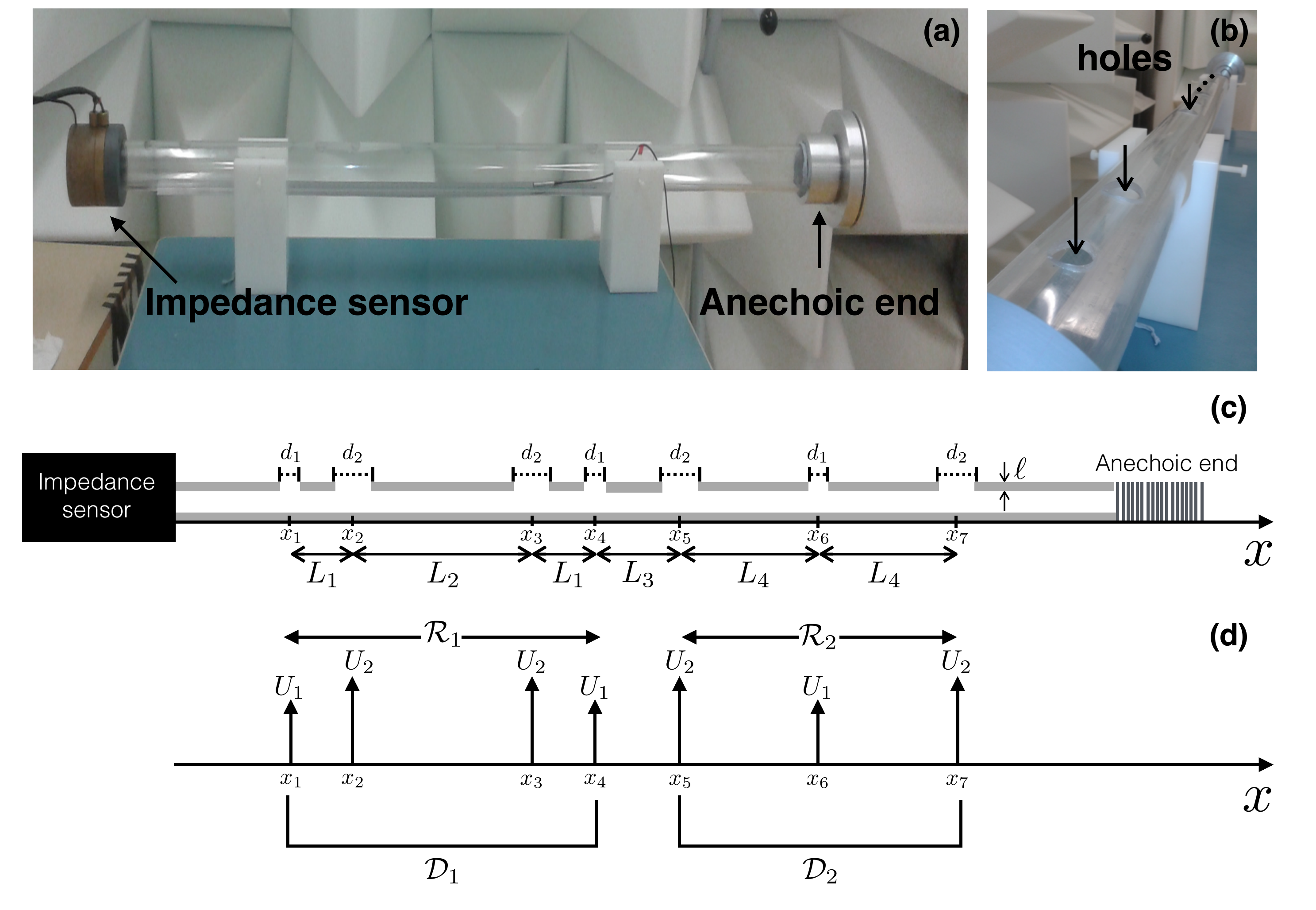}
\caption{ (Color online) (a),(b) Pictures of the experimental setup. (c) Schematic of the experimental setup. (d) $\delta$-barrier correspondence of the acoustic structure in the low frequency regime. $U_1$,$U_2$ correspond to the strength of the $\delta$-barrier potential and $D_1$, $D_2$ to the domains of the reflection symmetry decomposition of the waveguide..
}
\label{fig1}
\end{figure*}

In Fig. \ref{fig1} (a), we show a schematic of the aperiodic acoustic structure that was used in this work to experimentally characterize the transmission coefficient and the pressure fields of different frequencies. It consists of a cylindrical tube with section $S=\pi R^2$, perforated at the positions $x_n$ with holes of diameter $d_n= 2 r_n$ and length $\ell$, the thickness of the tube's wall. The sound source is a piezo-electric buzzer embedded in the impedance sensor (see \cite{ImpedanceSensor} for details) which is placed at one end of the main tube.
This sensor is used for the calculation of the transmission diagram and for the measurements of the pressure fields inside the lattice at prescribed frequencies. At the other end of the tube, a handmade termination is placed. This termination is a structure composed of a purely resistive metal tissue coupled with a cavity of tunable length (see appendix of \cite{Theocharis2014} for more details). By calculating the reflection coefficient of a tube, we found a reflection coefficient less than 5\% for the frequency range of $100-1000$ Hz. Thus, we consider this structure as an anechoic termination for our further analysis. 
A $1/2$ inch microphone (B\&K 4136), carefully calibrated, is used, placed after the last hole, for the calculation of the transmission coefficient and a $1/4$ inch microphone (GRAS Type 26AC) within the tube for field measurements. The frequency range of the applied signal is below the first cutoff frequency of the waveguide, $f_{c01}=4061$ Hz, and thus the propagation can be considered one-dimensional. 

\subsection{Wave Propagation} 

We consider the propagation of sound waves within an air-filled cylindrical tube of section $S$ perforated with cylindrical holes of section $s_n$ at positions $x_n$. If the geometric characteristics of the holes are much smaller than the wavelength of the propagating sound wave, i.e. $k r_n \ll 1$, the holes can be considered as point scatterers.
In a section of the tube between two consecutive point scatterers, the acoustic wave described by the pressure $p(x,t)$ is the solution of the wave equation:
\begin{equation}
\label{eq1}
\frac{\partial^2 p(x,t)}{\partial x^2}-\frac{1}{c^2}\frac{\partial^2 p(x,t)}{\partial t^2}=0
\end{equation}
where $c$ is the speed of sound in air. At each point scatterer (hole), the boundary conditions require the conservation of acoustic flow and continuity of acoustic pressure:
\begin{eqnarray}
\label{eq2a} v(x,t)|_{x_n^+}-v(x,t)|_{x_n^-} & = & -\frac{s_n}{S} v_h(x_n,t) \\
\label{eq2b} p(x,t)|_{x_n^+} & = & p(x,t)|_{x_n^-}
\end{eqnarray}
where $s_n$ is the cross section of the $n^{th}$ hole, $v(x,t)$ and $v_h(x_n,t)$ are the acoustic velocities inside the tube and inside the $n^{th}$ hole respectively. Taking the time derivative of Eq. (\ref{eq2a}) and using the $1D$ Euler equation, $\frac{\partial p(x,t)}{\partial x}=-\rho \frac{\partial v(x,t)}{\partial t}$,
we obtain 
\begin{equation}
\label{eq3a} \frac{\partial p(x,t)}{\partial x}\Bigg|_{x_n^+}-\frac{\partial p(x,t)}{\partial x}
\Bigg|_{x_n^-} = + \frac{\rho s_n}{S} \frac{\partial v_h(x_n,t)}{\partial t} 
\end{equation}
where $\rho$ is the air density. Eqs. \ref{eq1} and \ref{eq3a} lead to the inhomogeneous wave equation~\cite{Levine}:
\begin{eqnarray}
\label{eq4}
\frac{\partial^2 p(x,t)}{\partial x^2} & - &\frac{1}{c^2}\frac{\partial^2 p(x,t)}{\partial t^2} = 
\nonumber \\
& & \sum_{n}^{} \delta(x-x_n) \left( + \frac{\rho s_n}{S}\right) \frac{\partial v_h(x,t)}{\partial t}.
\end{eqnarray}
For plane waves the acoustic pressure  $p(x,t)$ and the velocity $v_h(x,t)$ are given by:
$$
p(x,t)=p(x)e^{-i \omega t} \quad \mbox{and} \quad v_h(x,t)=v_h(x)e^{-i \omega t},
$$
where $\omega= ck$ is the angular frequency. The amplitudes $p(x)$ and $v_h(x)$ at $x=x_n$ are related by the impedance relation $Z_n=p(x_n)/(s_n v_h(x_n))$.  At each connection between the waveguide and the hole, the wave impedance (which is modified by a change of cross sectional shape), the
acoustic  velocity  and therefore the derivative of the pressure $dp/dx$, are discontinuous functions. 
Using Eq. (\ref{eq4}), the pressure $p(x)$ is described by the following inhomogeneous Helmholtz equation: 
\begin{equation}
{\frac{d^2p(x)}{dx^2}}+k^2 p(x)= U(x)p(x),
\label{eq:etp}
\end{equation} 
where 
\begin{equation}
U(x)=\sum_{n}^{} \delta(x-x_n)U_n,
\label{eq:etp1}
\end{equation} 
and
\begin{equation}
U_n= -i k \frac{Z_c}{Z_n}.
\label{eq:sigma}
\end{equation} 
$Z_c=  \rho c/S$ is the characteristic impedance of the waveguide, $Z_n$ is the acoustic impedance of the n$^{th}$ hole at the point $x=x_n$ seen from the guide. In general the hole impedance $Z_n=R_n-iX_n$ is a complex function of the wavenumber, while the real part $R_n$ accounts for dissipation.

\subsection{Description of the losses}

The sound waves that propagate in the proposed acoustic structure are subject to two different kinds of losses. Propagation losses due to viscothermal effects on the wall, and point losses due to the scatterers. Viscothermal losses are taken into account by considering a complex expression for the wave number. In our case, we used the model of losses from Ref. [\onlinecite{Zwikker}], namely we replace the wave number and the characteristic impedance by the following expressions
\begin{eqnarray}
k=\frac{\omega}{c}(1+\frac{\beta}{s}(1+(\gamma -1)/\chi))
\label{KK}\\
Z_c = \frac{\rho c}{S}( 1+\frac{\beta}{s}(1-(\gamma -1)/\chi))
\label{ZK}
\end{eqnarray} 
by setting $s=r/\delta$ where $\delta=\sqrt{\frac{2\mu}{\rho \omega}}$ is the viscous boundary layer thickness, $\mu$ being the viscosity of air, $\chi=\sqrt{P_r}$ with $P_r$ the Prandtl number, $\beta=(1-i)/\sqrt{2}$, $\gamma$ the heat capacity ratio of air and $r$ the radius of the considered tube.

Scatterer losses are taking into account in the real part of the acoustic impedance of the holes. They come from viscothermal losses at the boundaries of the holes wall and radiation losses to the outer environment. Thus, the impedance of the hole can be expressed as:
\begin{equation}
Z_n= R_n -i Z_h \tan (k (\ell + \ell_{i,n} + \ell_{o,n})),
\end{equation}
where $R_n= \frac{\rho c}{2 s_n} (k r_n)^2$ for $k r_n \ll 1$ \cite{Nomura}, $Z_h=  \rho c/s_n $.
$\ell_{i,n}$ and $\ell_{o,n}$ are length corrections due to the radiation inside the principal waveguide \cite{Dubos} and to the outer environment \cite{Dalmont01} respectively (see appendix \ref{sec:AppA} for detailed calculations).

Note that for the frequency range of our interest, $k (\ell + \ell_{i,n} + \ell_{o,n}) \ll 1$. Thus, ignoring the losses ($R_n=0$) and after Taylor expansion of the tangential function, $Z_n\approx - i Z_h\times k (\ell + \ell_{i,n} + \ell_{o,n})$. In that case, the potential $U(x)$ becomes a series of $\delta$ barriers with strength: 
\begin{equation}
U_n=\frac{s_n}{S(\ell + \ell_{i,n} + \ell_{o,n})}.
\label{U_n_d}
\end{equation}

\section{Acoustic Waveguide with Complete Local Symmetry}\label{setup-design}

\subsection{Construction rules for perfect transmission resonances}

The transmission spectra and the control of their properties in aperiodic, inhomogeneous media is a field of intense study 
among a large class of wave scattering systems~\cite{Ren2015,Nomata2007,Fung2005}. The design of our experimental setup has been based on the (lossless) construction rules~\cite{Kalozoumis2013a,Kalozoumis2013b} for perfect transmission resonances, which provide the technique for the design of setups which are transparent at preselected frequencies. 
 
In general, isolated perfect transmission resonances are expected to occur in reflection symmetric {\it lossless} systems. On the contrary, in aperiodic, nonsymmetric systems, the transmission spectrum is usually characterized by $T=|t|^2<1$, where t is the transmission coefficient. Nevertheless, the possible complete
decompositions of an aperiodic CLS setup into locally reflection symmetric potential parts, has a major impact on its transmission 
properties and particularly on the emergence of perfect transmission resonances. The correspondence between perfect transmission 
and local symmetries has been demonstrated for quantum mechanical~\cite{Kalozoumis2013a} and photonic~\cite{Kalozoumis2013b} systems.
Recent results~\cite{Huang2001,Nava2009,Zhukovsky2010}, reporting on perfect transmitting resonances in globally asymmetric setups,
are unambiguously explained within the framework of local symmetries.

Perfect transmission resonances can be distinguished into the following two classes according to the spatial profile of $|p(x)|$ along the device~\cite{Kalozoumis2013b}:

$s$-PTRs: Symmetric PTRs are in one-to-one correspondence with the multitude of the different possible decompositions of the setup into reflection symmetric potential parts~\cite{Kalozoumis2013a}. In this case, the spatial profile of the wave field follows the local reflection symmetries of the selected decomposition. In order to construct $N$ $s$-PTRs at $N$ preselected frequencies $f_{i},~i=1,..,.N$, $N$ different such decompositions should be identified. Following the construction principle developed in~\cite{Kalozoumis2013a,Kalozoumis2013b}, we can find the relevant $\delta$-barrier parameters (strengths $U_n$ and distances $L_n$) which allow for the specific perfect transmission properties.  Subsequently, these values are used to determine the geometric characteristics of the setup, e.g. the radius of the  holes which play the role of the point scatterers, and their distances.

$a$-PTRs: The procedure to construct an asymmetric PTR requires to divide the setup into two arbitrary (not necessarily reflection symmetric) parts $\mathcal{R}_1$ and $\mathcal{R}_2$. The first step is to compute the transmission spectra for $\mathcal{R}_1$ and $\mathcal{R}_2$ independently and find the frequencies of their intersections. At these frequencies, $a$-PTRs are possible to occur by finding the suitable distance between $\mathcal{R}_1$ and $\mathcal{R}_2$, such that the relevant conditions shown in Ref.~\cite{Zhukovsky2010} are satisfied. In this case, the spatial profile of the wave field does not follow any (local) symmetry within the corresponding setup.

\subsection{Design based on PTR construction rules and minimization of the losses}

Based on the aforementioned construction technique we theoretically design a setup which -in the lossless case- manifests both kinds of PTRs. Subsequently, the emerging geometric characteristics ($\delta$-barrier strengths and their distances) are implemented in the fabrication of the acoustic waveguide. In order that the experimental procedure yields satisfactory results, attenuation should be taken into account and controlled. Therefore, to minimize the propagation losses, the waveguide
has to be relatively short, of the order of one meter.  Radiation losses, on the other hand, are enhanced as the number of the scatterers (holes) increases, due to the acoustic radiation to the outer environment. An optimal choice which satisfies these experimental objectives is schematically illustrated in Fig.~\ref{fig1} and  
described below.  Nevertheless, the emerging (acoustic) structure can be considered as a weakly lossy system. Let us note that the periodic/aperiodic distribution of strong losses can lead to interesting phenomena like formation of antibandgap~\cite{Staliunas2009,Cebrecos2014}.

The final aperiodic CLS setup is comprised of seven scatterers of two different potential strengths $U_{1}=7.8886$ and $U_{2}=12.3414$ corresponding to holes of radii $r_{1}=0.009~m$ and $r_{2}=0.012~m$, respectively (for the calculation of potential strengths, we use Eq. (\ref{U_n_d})).
Our choice to use a limited number of scatterers in order to minimize losses, results in the identification of a single decomposition in reflection symmetric parts (defining the domains $\mathcal{D}_{1}$,~$\mathcal{D}_{2}$), which are depicted by the solid arcs in Fig.~\ref{fig1}(d). Accordingly, the correspondence between different decompositions in reflection symmetric parts and $s$-PTRs dictates that from this setup only one $s$-PTR can emerge.  
Selecting the $s$-PTR frequency to be $f_{s}=450$~Hz and implementing the corresponding construction principle, we find that the distances between the scatterers are $L_{1}=0.07$~m, $L_{2}=0.2671$~m and $L_{4}=0.1925$~m.

Note that the $s$-PTR is not affected by the variation of the length $L_{3}$ separating the two reflection symmetric parts of the decomposition. This free parameter will be utilized for the construction of an additional $a$-PTR in the following manner: By identifying two parts of the device i.e. $\mathcal{R}_{1},~\mathcal{R}_{2}$, we calculate independently their transmission spectra and seek for frequencies with common transmission coefficient. For the case studied here this condition is fulfilled at $f_{a}=820$~Hz. Fixing this frequency and varying the distance $L_{3}$ we find that the $a$-PTR occurs when $L_{3}=0.1047$~m.

\subsection{Invariants of broken discrete symmetries in systems with losses}\label{invariants}

In~\cite{Kalozoumis2014a}, it has been shown that in the case of scattering systems characterized by a globally broken discrete symmetry, a nonvanishing non-local, spatially invariant current $Q$ exists that characterizes the symmetry breaking. This analysis, however, concerned lossless idealized systems. In experimental, realistic setups losses are inevitable and their theoretical account is necessary to accurately describe or predict the expected outcome. In the following, we develop a formalism which systematically describes the breaking of discrete symmetries in systems with attenuation and we elucidate how the quantity $Q$ is affected.

Losses may emerge either during the propagation or locally at the scatterer points due to acoustic radiation to the outer environment and viscothermal effects.
Propagation losses for structures of small length can safely be neglected when compared with the attenuation emerging from the local scatterers. This means that the impedance of the hole becomes a complex function and therefore the potential obtains the following form:
\begin{equation}
\label{compl_pot} U(x)=U_{R}(x)+iU_{I}(x).
\end{equation}

The condition that allows the applicability of the theory is that both the real and the imaginary parts of the potential $U(x)$ obey the linear symmetry transform $U(x)=U(F(x))$ for  $F(x)\equiv \bar{x}=\sigma x + \eta$. Depending on the value of the parameters $\sigma$ and $\eta$, $F(x)$ describes a reflection at a point $\alpha$ ($\sigma=-1,~\eta=2\alpha$) or a translation by a length $\beta$ ($\sigma=1,~\eta=\beta$). The following analysis holds for both symmetry transformations whereas in this work we focus exclusively on reflection symmetric potentials, 
\begin{eqnarray}
\label{lossy_potential_real_imag} U_{R}(x)&=&U_{R}(\bar{x}), \nonumber \\
 U_{I}(x)&=&U_{I}(\bar{x}).
\end{eqnarray}
The potential in Eq.~(\ref{eq:etp1}) which is used in this work satisfies both conditions. 

To obtain $Q$, we multiply Eq.~(\ref{eq:etp}) with $p(\bar{x})$ and its parity transform with $p(x)$. Then we subtract the resulting equations and due to the symmetry expressed in Eq.~(\ref{lossy_potential_real_imag}) the potential terms cancel and the resulting expression can be written as a total derivative. 
This, in turn, leads to the intriguing result that $Q$ remains spatially invariant even in the presence of attenuation,
\begin{equation}
\label{Q}
Q=\frac{1}{2i}\left[\sigma p(x)p'(\bar{x})- p(\bar{x}) p'(x) \right] = const,
\end{equation}
indicating the strength of the constraints which are imposed by the symmetry. 
Dividing Eq.~(\ref{Q}) by $p^{2}(x)$ we obtain,
\begin{equation}
\label{lossy_bl_par_1} \frac{Q}{p^{2}(x)}=\frac{1}{2i}\left(\frac{\sigma  p(x)
p'(\bar{x})- p(\bar{x}) p'(x)}{p^{2}(x)}\right)
\end{equation}
or
\begin{equation}
\label{lossy_bl_par_2} \frac{Q}{p^{2}(x)}=\frac{1}{2i}\left(\frac{p(\bar{x})}{p(x)}\right)^{\prime}.
\end{equation}
By integration we find
\begin{equation}
\label{lossy_bl_par_3} p(\bar{x})=cp(x)-2iQ~p(x)\int_{x}^{\zeta}\frac{1}{p^{2}(\xi)}d\xi,
\end{equation}
where $c=p(\bar{\zeta})/p(\zeta)$. 
Equation~(\ref{lossy_bl_par_3}) constitutes a mapping for the pressure field, linking $p(x)$ from the source part to the target part of the symmetry domain, which are related via the symmetry transformation $x \to \bar{x}$. Obviously, this transform maps $[x,\zeta]$ to $[\bar{x},\bar{\zeta}]$. This property is of fundamental significance, since it generalizes the well-known Bloch and parity theorems, not only for systems with globally broken reflection symmetry~\cite{Kalozoumis2014a}, but also when they sustain losses. The latter renders the constancy of $Q$ directly observable in realistic and experimentally realizable lossy wave scattering setups. The experimental verification of this result will be presented in the following section, focusing on the case of reflection symmetry.

\begin{figure}[t!]
\begin{center}
\includegraphics[width=1.\columnwidth]{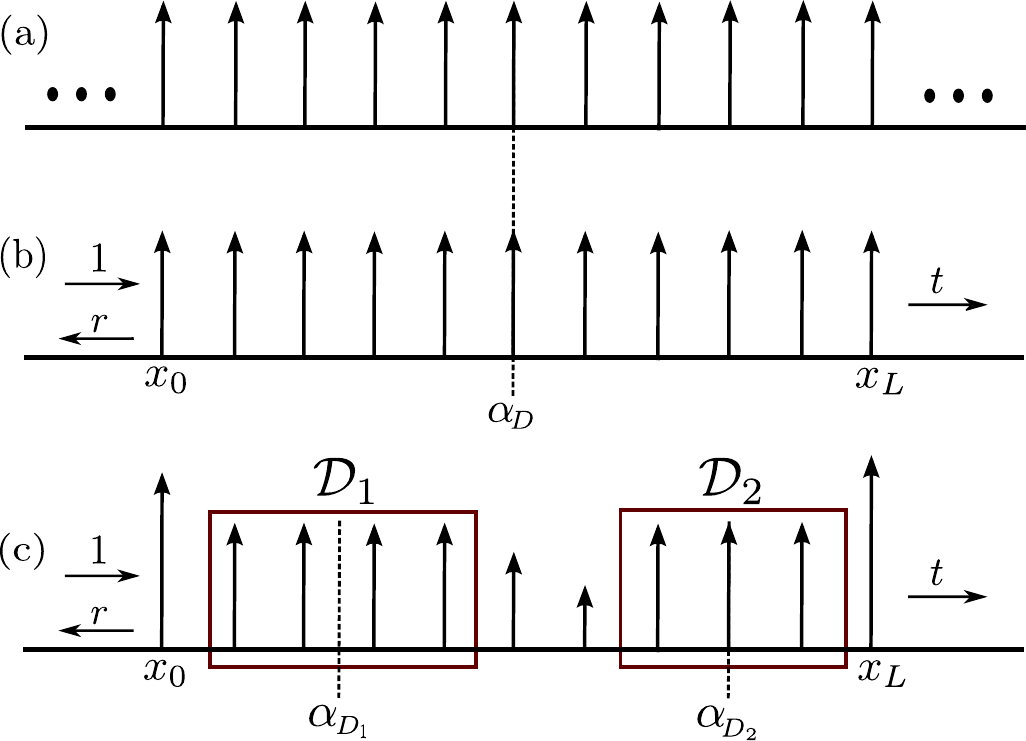}
\end{center}
\vspace{.2cm}
\caption{\label{fig2} (Color online) (a) Schematic of an infinite system with global reflection and translation symmetry. (b) The global reflection symmetry is broken due to the asymmetric asymptotic conditions. However, the setup is still reflection symmetric.  (c) The reflection symmetry of the setup is explicitly broken under the presence of two defects. Nevertheless, it is retained locally in the restricted spatial domains $\mathcal{D}_{1},~\mathcal{D}_{2}$. }
\end{figure}

The importance of the invariant current $Q$ can be perceived within the systematic description of symmetry breaking. In a globally
symmetric system the corresponding symmetry is satisfied $\forall~ x \in \mathbb{R}$. A reflection (or translation) symmetric system extending infinitely, as the one illustrated in Fig.~\ref{fig2} (a) belongs to this class. A vanishing $Q$ value indicates this case. In scattering systems, however, the global symmetries appear broken and $Q$ is the key tool to track the mechanism of the symmetry breaking. Here it is possible that the structure possesses
a global symmetry, whereas this symmetry is globally broken due to the imposed asymptotic conditions. In this case, which is illustrated in Fig.~\ref{fig2} (b),  the invariance of $Q$ is retained globally, in the same sense that the energy current is constant in the lossless case.
For the typical scattering asymptotic conditions,
\begin{eqnarray}
\label{asympt_cond} p_{ \textrm{L}}(x)&=&e^{ik(x-x_0)}+re^{-ik(x-x_0)},~~ x-x_0< 0 \nonumber \\
                    p_{ \textrm{R}}(x)&=&te^{ik(x-x_L)},~~~~~~~~~~~~~~~~~~~ x-x_L>0
\end{eqnarray}
with $p_{L}$ and $p_{R}$ being the pressure field on the left and right side of the setup, respectively, $L=x_L-x_0$ being the length of the setup (see Fig.~\ref{fig2} (b)) and $r$ and $t$ corresponding to the  pressure field reflection and transmission amplitudes, it is straightforward to show that
\begin{eqnarray}
\label{Q_magn_refl} |Q|&=&kT^{\frac{1}{2}},~~~~~~~~~~~\textrm{for reflection symmetry} \\
\label{Q_magn_trans} |Q|&=&k(RT)^{\frac{1}{2}},~~~~~~\textrm{for translation symmetry}
\end{eqnarray}
with $T=|t|^{2}$ and $R=|r|^{2}$ being the pressure field transmission and reflection coefficients for the lossy system, respectively. Particularly for the reflection symmetry transformation, Eq.~(\ref{Q_magn_refl}) has a similar form as the energy current $J=kT_{\textrm{lossless}}$ for the lossless analogue, differing only in the power of the transmission coefficient. It is worth mentioning that despite the fact that in a system with attenuation the energy current $J$ ceases to be constant, the symmetry induces a constant current regardless of the losses. 

A more complete breaking of the underlying symmetry however takes place if we do not only introduce an asymmetry of the scattering asymptotic conditions. In this case the potential itself globally does not obey the corresponding symmetry transformation but the symmetry is present locally i.e., in some spatially limited domains. An indicative such case is shown in Fig.~\ref{fig2} (c) where the reflection symmetry is globally broken due to the presence of defects, nonetheless it is locally retained  in the domains $\mathcal{D}_{1},~\mathcal{D}_{2}$. Whereas here there is no spatially invariant current Q for the total extent of the device, there
exist two different currents $Q_1,~Q_2$ spatially invariant within the symmetry domains $\mathcal{D}_{1},~\mathcal{D}_{2}$, respectively. Breaking further the symmetry would lead to smaller regions of constant $Q$s. In the extreme case of a complete breaking of the corresponding symmetry i.e. strongly disordered systems without any persisting local symmetry, Q ceases being constant. In this sense, the landscape of $Q$ provides a systematic tracking of the breaking and the corresponding remnants of a symmetry.

\section{Results}\label{results}

\subsection{Transmission spectrum}

\begin{figure}[t!]
\begin{center}
\includegraphics[width=.98\columnwidth]{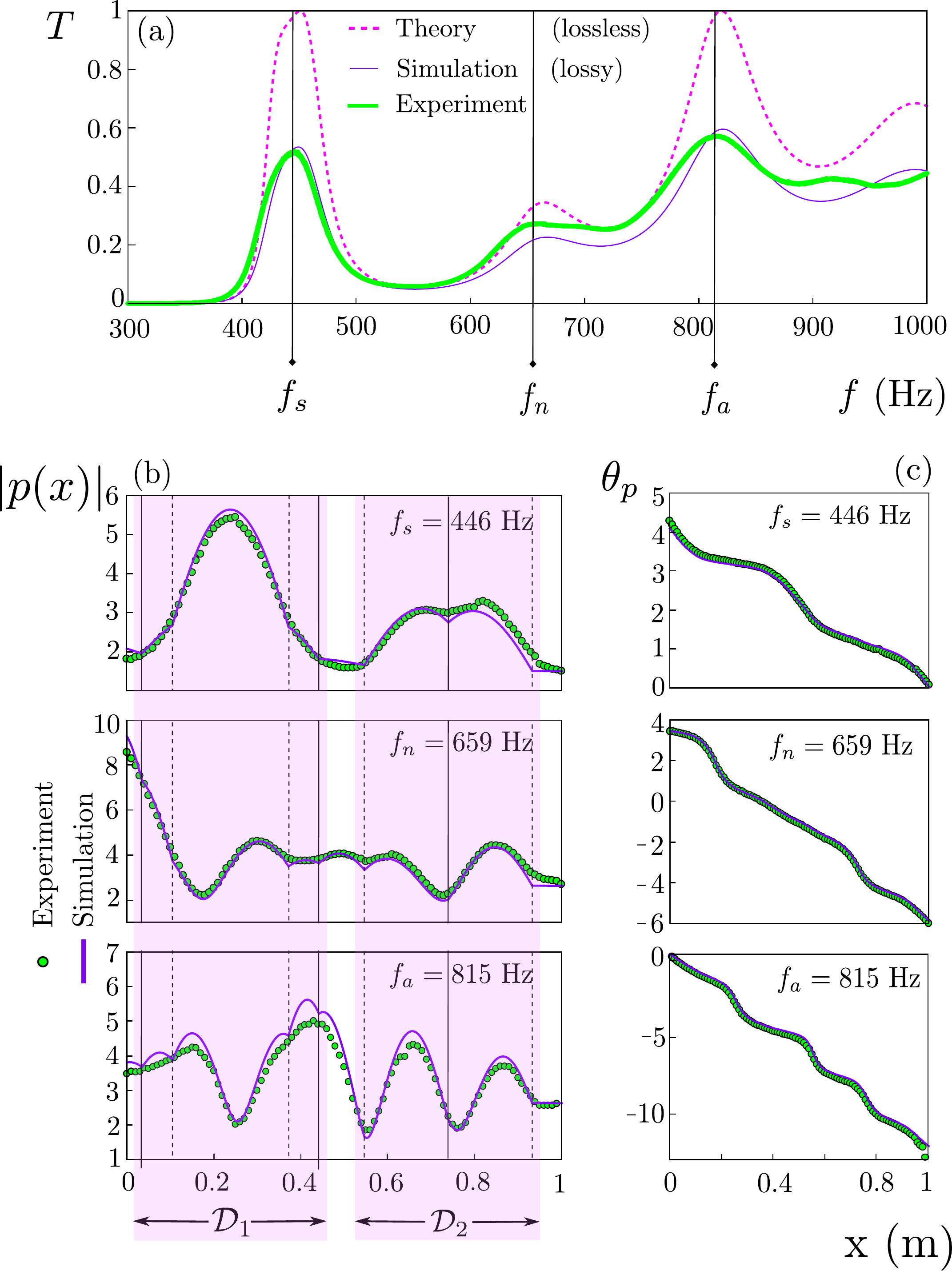}
\end{center}
\vspace{.2cm}
\caption{\label{fig3} (Color online) (a) Transmission spectrum of the aperiodic waveguide: lossless theory (magenta short dash curve), theory with losses (purple long dash curve)
and experiment (green solid curve). The vertical (solid) lines indicate the frequencies we focus on: $s$-PTR frequency ($f_s$), $a$-PTR frequency ($f_a$) and non-PTR ($f_{n}$). (b) Pressure field magnitude $|p(x)|$ across the waveguide at the frequencies $f_{s},f_{a},f_{n}$ marked in (a). The solid and dashed vertical lines show the positions of the centers of the holes with two different radii and thus different scattering strength. The shaded areas depict the two domains of reflection symmetry ($\mathcal{D}_{1},~\mathcal{D}_{2}$) to which the setup can be completely decomposed. 
(c) Phase of the pressure field $\theta_{p}$ across the waveguide for the three selected frequencies $f_s$, $f_n$ and $f_a$.}
\end{figure}

Having determined the geometric characteristics of the waveguide, we study experimentally the local symmetry induced characteristics at the  PTR frequencies. Figure~\ref{fig3} shows the comparison between the theoretical lossless, the theoretical lossy and the experimental transmission spectra. The theoretical transmission, both for the lossless and lossy case, are obtained by the transmission matrix method (see Appendix \ref{sec:AppB} for details).
The experimental transmission spectrum is obtained by calculating the transmission matrix  elements following the procedure described in \cite{Theocharis2014} and using an anechoic termination. In addition, due to the asymmetry of the structure, two different measurements (source at the right of the structure, anechoic end at the left for the first measurement and the reverse position for the second) are performed. 

The experimental transmission spectrum is in very good agreement with the lossy theory. Small discrepancies between theory and experiment can be
attributed to the reflections at the termination end, to a non-perfect calibration of the microphone and to unavoidable leaks of the system. The obvious effect of losses is the decrease of the PTR peaks at a percentage of $40-50\%$. Note also that in the lossy case the peaks corresponding to PTRs are slightly shifted and appear at frequencies  $f_{s}=446$~Hz and $f_{a}=815$~Hz instead of the theoretically predicted
values of $f_{s}=450$~Hz and $f_{a}=820$~Hz, respectively.  

\subsection{Pressure profiles}

Figures~\ref{fig3} (b)-(c) illustrate the magnitude of the pressure field, $|p(x)|$, and its corresponding phase $\theta_{p}$ for $f_{s}$, $f_{a}$ and $f_{n}=659$~Hz as indicated in Fig.~\ref{fig3} (a) by the dashed lines, where $f_{n}$ corresponds to a non-PTR state.
Experimental measurements of the magnitude and the phase of  the pressure field at these frequencies  were performed by a $1/4$ inch microphone (GRAS Type 26AC) placed within the tube and positioned at lattice points of $dx=1$~cm, see Fig. \ref{fig1}(a).

The theoretically anticipated results (with losses incorporated) are obtained by the transmission matrix (see Appendix \ref{sec:AppB} for detailed calculations) using anechoic termination conditions, namely an impedance termination equal to the acoustic characteristic impedance of the tube as boundary condition. Theoretical and experimental results, both for $|p(x)|$ and for $\theta_{p}$ are in very good agreement. We also observe that the magnitude of the pressure field at the $s$-PTR shown in Fig.~\ref{fig3}(b) follows the symmetries of the mirror symmetric domains which constitute the corresponding decomposition, as expected from the lossless theory~\cite{Kalozoumis2014b}.  The small deviations from the exact symmetry are attributed to the losses sustained by the system and to the small back propagation effects due to imperfections of the anechoic end. 

The spatial profiles of the magnitude of the pressure field indicate an additional interesting property of the CLS class of devices, which concerns the localization of the field. Particularly, Fig.~\ref{fig3} (b) indicates that at the $s$-PTR frequency the localization of the pressure field is enhanced in the first mirror symmetric domain compared to the second. Similar results have been theoretically observed in~\cite{Kalozoumis2013a,Zhukovsky2010}, evincing the potential usefulness of this class of structures for the control of localization.  

\subsection{Observation of the symmetry induced invariant currents}
Let us experimentally verify the presence of the invariant current $Q$. For reasons of consistency we choose for the analysis the same selected frequencies,  $f_{s}$, $f_{a}$ and $f_{n}$ to calculate the current $Q$. We follow two different procedures for this demonstration.
\subsubsection{Plane wave fitting method}
This method assumes an one-dimensional sound propagation. Using the experimental pressure field data, we perform a fitting procedure with a plane wave function of complex wavenumber $k=k_{R}-ik_{I}$ given by the expression 
\begin{eqnarray}
f(x)=\vert A \vert e^{-k_{I}x} \cos(\frac{\epsilon \pi}{2}-(\theta_A + k_{R} x)) &  \nonumber \\ +\vert B \vert e^{k_{I}x} \cos(\frac{\epsilon \pi}{2}-(\theta_B - k_{R} x)).
\label{plane_wave}
\end{eqnarray} 
We fit simultaneously the real and imaginary parts of the experimental pressure field using $\vert A \vert$, $\vert B \vert$, $\theta_A$, $\theta_B$, $k_{R}$ and $k_{I}$ as shared parameters for the two datasets (real, imaginary part). When the parameter $\epsilon$ is zero, the fitting function corresponds to the real part of the field, whereas for $\epsilon=1$ Eq.~(\ref{plane_wave}) corresponds to the imaginary part. Then, keeping the complex wave number fixed, we find the plane wave parameters  $A$  and $B$. 

\begin{figure*}
\begin{center}
\includegraphics[height=8cm,width=0.9\textwidth]{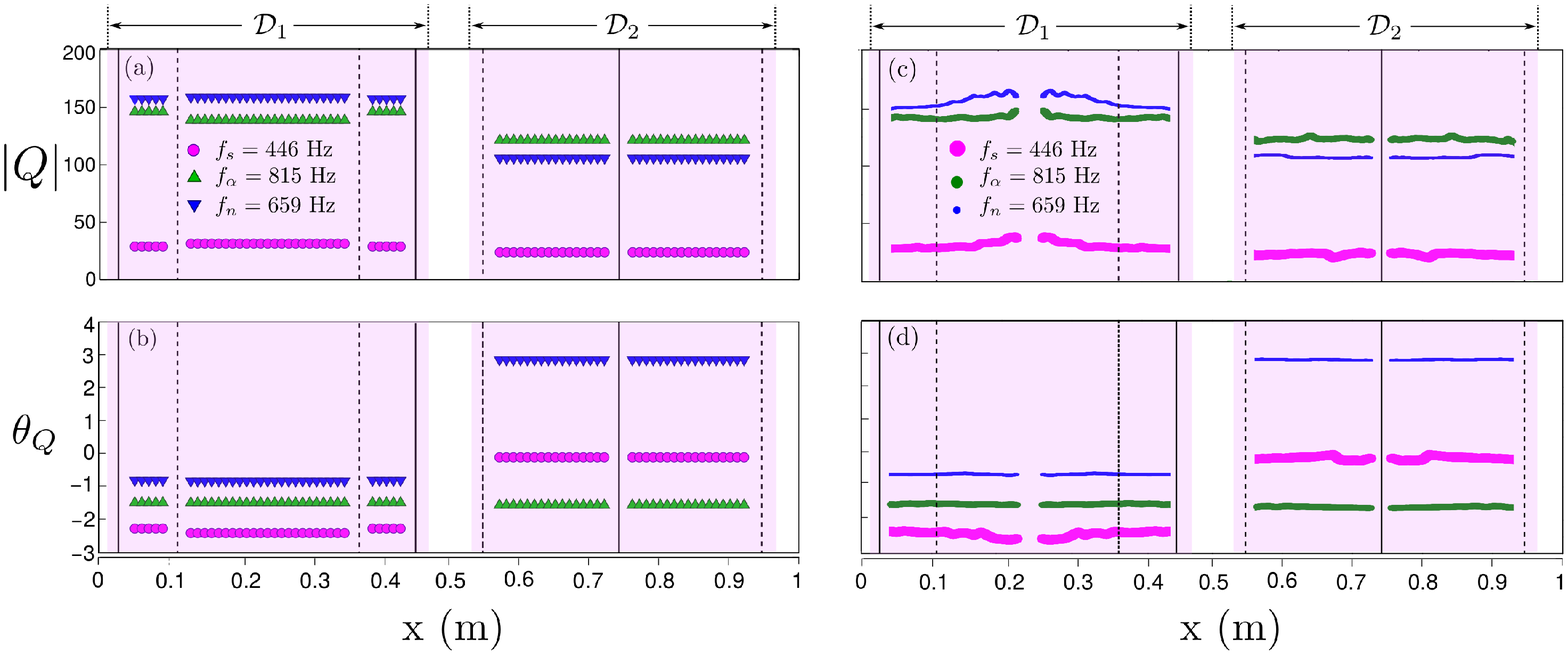}
\end{center}
\vspace{.2cm}
\caption{\label{fig4} (Color online) Schematic of the setup, illustrating the spatial profile of $|Q|$ and $\theta_{Q}$. The solid and dashed lines correspond to the positions and strengths $U_{1}$ and $U_{2}$ of the point scatterers, respectively. The coloured areas delimit the domains $\mathcal{D}_{1},~\mathcal{D}_{2}$ from the start of the first hole to the end of the final hole. The magnitude and phase of the invariant current $Q$ are calculated for the selected $s$-PTR ($f_{s}=446$ Hz), $a$-PTR ($f_{a}=815$ Hz) and non-PTR ($f_{n}=659$ Hz) frequencies. (a) and (b) show the magnitude and the phase of $Q$ calculated with the method of fitting the experimental data in each potential free area with the proper plane wave. Accordingly, (c) and (d) show $|Q|$ and $\theta_{Q}$ calculated using the mapping relation  (Eq.~(\ref{lossy_bl_par_3})). }
\end{figure*}

However, around the holes, the scattered waves have an evanescent part with 3D characteristics which locally alter the plane wave description. Thus, we ignore the experimental points residing in the areas which extend one millimetre on either side of each hole and where a deviation from the plane wave description occurs. The remaining experimental points comprise five regions -three in the domain $\mathcal{D}_{1}$ and two in the domain $\mathcal{D}_{2}$- where the plane wave approximation is accurate. Thus, we perform  a separate fitting procedure in each of these five regions.
The invariant currents $Q$ are determined by using the results of the fitting procedure in Eq.~(\ref{Q}).
Figures~\ref{fig4} (a), (b) illustrate $|Q|$ and the respective phase $\theta_{Q}$ for the three selected frequencies.

\subsubsection{Mapping method}

The alternative way to calculate $Q$ uses directly the experimental data, without excluding any regions or using fitting procedures. Having measured the pressure field in the total extent of the acoustic structure, we can directly compute $Q$ via the mapping relation,
\begin{equation}
\label{Q_2} Q=\frac{p(x)-p(2\alpha-x)}{2i~p(x)\int_{x}^{\alpha}\frac{1}{p^{2}(\xi)}d\xi},
\end{equation}
which results from Eq.~(\ref{lossy_bl_par_3}). Figures~\ref{fig4} (c), (d) illustrate the corresponding results for $|Q|$ and $\theta_{Q}$, as emerged from the latter. The integral in Eq.~(\ref{lossy_bl_par_3}) becomes zero at the mirror axis position of each domain and the expression becomes indeterminate. A more accurate approach of this vicinity would require more experimental data points. Therefore we have ignored a small area around the domain center. This also causes the observed increasing trend as the center of the domains is approached.  The small fluctuations of the values of $|Q|$ and $\theta_{Q}$ are inevitable since the experimental points deviate from the exact plane wave which would be the solution of Eq.~(\ref{eq:etp}).

\subsubsection{Remarks}

Comparing the results of the two approaches it is obvious that they are in very good agreement and the constancy of $Q$ in the symmetry domains becomes evident, verifying the theoretical prediction. In particular, since our structure is composed of two local symmetry subdomains $\mathcal{D}_{1},~\mathcal{D}_{2}$, according to the lossless theory~\cite{Kalozoumis2013b} the magnitude of the invariant currents $|Q|$ in each domain at the PTR frequencies should be the same. At any other non-PTR frequency, the magnitudes of the invariant currents $Q$ is constant in each domain of the selected decomposition, but they differ from each other. 

In the first place,  both for theory and experiment (see Figs. \ref{fig4} (a) and (c)), we observe that in the presence of attenuation, $|Q|$ is different at the two decomposition domains even at the PTR frequencies, thereby
deviating from the lossless theory. Using the theoretical transmission matrix method and tuning the total amount of losses from zero up to the expected value, we observe that as attenuation increases, the larger the difference between $|Q|$ in $\mathcal{D}_{1}$ and $\mathcal{D}_{2}$, becomes. On the other hand, for a given amount of losses which is kept fixed, by changing the frequency, we observe that this difference is minimized at the $s$-PTR frequency, being in accordance to the lossless theory \cite{Kalozoumis2013a}.

An additional feature which is observed in the plane wave fitting method and is worth-mentioning, concerns the small gaps which appear in the experimental data within the domain $\mathcal{D}_{1}$ (resp. $\mathcal{D}_{2}$) and particularly between the first (or third for $\mathcal{D}_{1}$) region and the second. This can be explained by the  three-dimensional nature of the problem in the vicinity of the holes. This feature alters the solution from one region to the other, leading to disconnected plane wave solutions for each one of the three regions, instead of the exact, continuous plane wave for the whole first domain, as in the case of the theoretical results.
This difference in the plane wave solutions is therefore translated into a gap for $|Q|$ and $\Theta_{Q}$. 

Concluding this subsection, we point out that a complete constancy of $Q$ would be feasible if the pressure field in every symmetry domain was a perfect plane wave, satisfying the 1D Helmholtz equation. The deviation of the experimental curve from the exact
functional form of the plane wave results in a gap in domain $\mathcal{D}_{1}$ when the fitting procedure is followed or in
small fluctuations when the calculation is performed via Eq.~(\ref{lossy_bl_par_3}).

\subsection{Field mapping between symmetry related areas}

\begin{figure*}[htbp!]
\vspace{1cm}
\begin{center}
\includegraphics[width=0.8\textwidth]{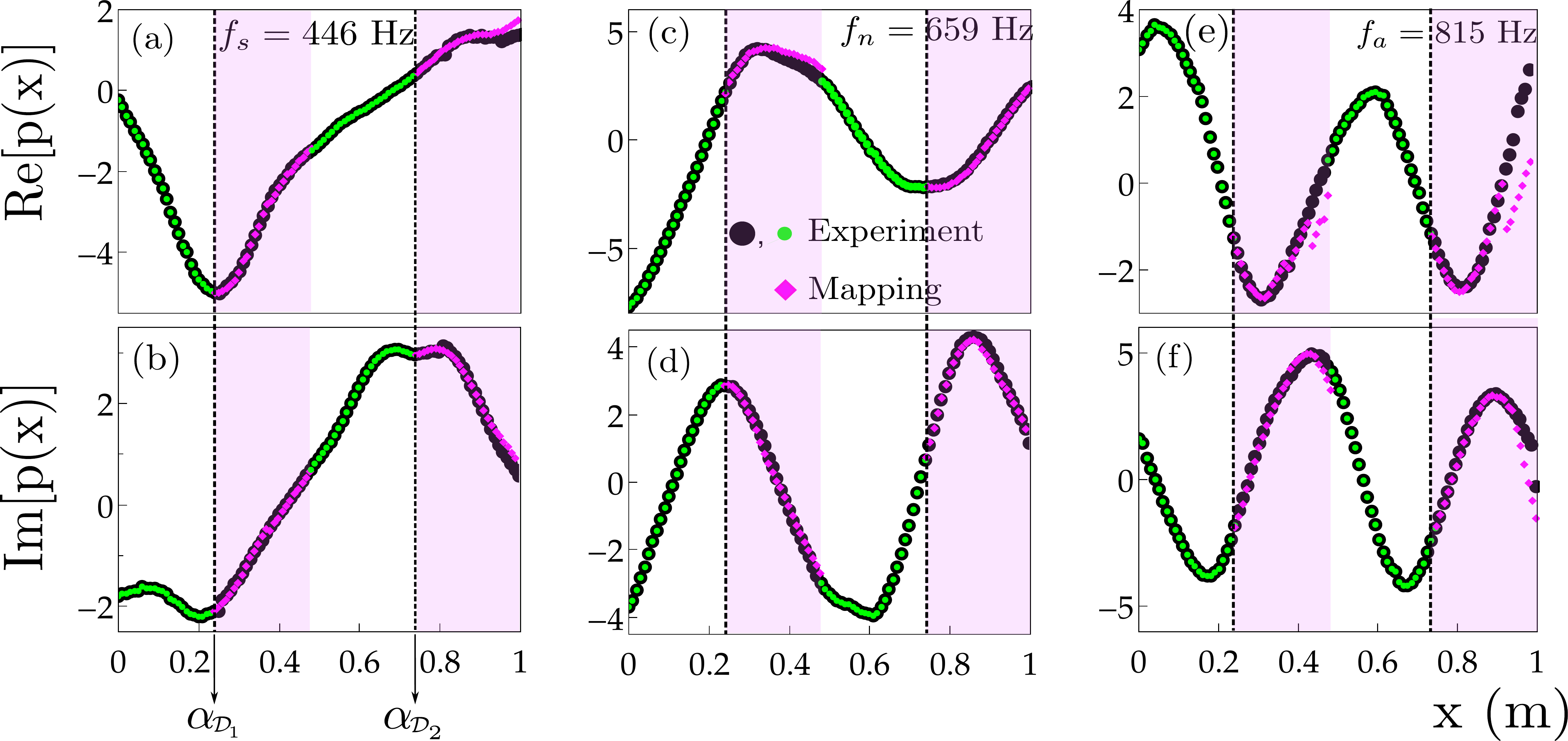}
\end{center}
\vspace{.2cm}
\caption{\label{fig5} (Color online) Real and imaginary parts of the pressure field $p(x)$ for the selected $s$-PTR ($f_{s}=446$ Hz), $a$-PTR ($f_{a}=815$ Hz) and non-PTR ($f_{n}=659$ Hz) frequencies. The black (large) circles correspond to the experimental field's spatial profile. The green small circles (always coinciding with the black larger circles) also correspond to the experimental data and are used to indicate the experimental data points (source domain in white) which are mapped through Eq.~(\ref{lossy_bl_par_3}) to the magenta diamonds (target domain in magenta).}
\end{figure*} 

In Ref.~\cite{Kalozoumis2014a}, it was demonstrated how the spatially invariant current $Q$ constitutes the key tool for tracking the symmetry breaking procedure in lossless, aperiodic systems. Here, having already identified the constancy of $Q$ in the presence of losses, we experimentally verify the generalization of the parity theorem in the case of complete local reflection symmetry in the presence of losses. To this aim, we utilize the aperiodic setup considered above (see Fig.~\ref{fig1}) and use the measured $Q$s and pressure fields for the frequencies $f_{s},~f_{a}$ and $f_{n}$. The mapping between the symmetry related regions is provided by Eq.~(\ref{lossy_bl_par_3}).

In Fig.~\ref{fig5} the real and imaginary parts of the pressure field are illustrated for the selected $s$-PTR ($f_{s}=446$ Hz), $a$-PTR ($f_{a}=815$ Hz) and non-PTR ($f_{n}=659$ Hz) frequencies. To investigate the validity of the above mapping relation expressed in Eq.~(\ref{lossy_bl_par_3}), we employ as input parameters the experimental pressure field at the left side of each mirror symmetry axis, represented by the green (small) circles and the $Q$ values obtained by the fitting procedure, presented in the previous section (Fig.~\ref{fig4} (a), (b)). Note that we use the $Q$ values obtained by the fitting method in order to have an independent mapping result.
The resulting mapped points, denoted by the magenta diamonds, are in very good agreement with the experimental data.

\section{Conclusions}\label{conclusions}

This is the first experimental implementation and verification of the concept of local symmetries and the construction principle proposed in Refs.~\cite{Kalozoumis2013a,Kalozoumis2013b} for aperiodic acoustic waveguides with prescribed perfect transmission resonances.
The experimental observation reveals the universal character of this concept, which has been theoretically predicted to occur in quantum mechanical aperiodic structures~\cite{Kalozoumis2013a} and photonic multilayered structures~\cite{Kalozoumis2013b}, but is now experimentally verified in a completely different, namely acoustic, system. In particular, we extend the theory by incorporating losses and ascertain that the symmetry induced invariant current $Q$ preserves its spatial constancy in the presence of attenuation. The fact that broken discrete symmetries, retained at spatially limited domains, induce invariant currents, has been experimentally verified here for the case of locally reflection symmetric setups. In turn, we showed how $Q$ generalizes the Bloch and parity theorems for scattering systems with broken global translation and reflection symmetry, respectively, by defining a mapping of the pressure field between the symmetry related regions. Subsequently, we experimentally verified the validity of this generalization for the case of reflection symmetries. It has been demonstrated how $Q$ constitutes the key-tool towards a systematic description of discrete symmetry breaking. In this context, we showed that in a globally reflection symmetric setup which sustains losses, $|Q|$ provides the lossy analogue to the energy current J of the lossless case. Finally, the construction procedure for the design of acoustic waveguides with prescribed perfect transmission properties and the subsequent experimental implementation reveals how the class of CLS materials could pave the way towards the design of materials with very specialized filtering capabilities.

\appendix
\section{Length corrections due to the hole}
\label{sec:AppA}

For $k r_n \ll 1$, Refs. \cite{Dubos} and \cite{Dalmont01} give the following expressions for, respectively, the length correction 
due to radiation inside the principal waveguide and due to radiation to outer environment, respectively:
\begin{equation}
\ell_{i,n} = 0.82 (1-0.235 \epsilon_n - 1.32 \epsilon_n^2 + 1.54 \epsilon_n^3 - 0.86 \epsilon_n^4)r_n
\end{equation}
and
\begin{equation}
\ell_{o,n}=(0.82-0.47^{(0.8)}\epsilon_n) r_n,
\end{equation}
with $\epsilon_n=r_n/R$.

\section{Transmission Matrix Method}
\label{sec:AppB}

Applying the transmission matrix method, the propagation through a finite system made of $N_h$ holes can be expressed as follow \cite{Theocharis2014}
\begin{equation}
\left(
\begin{array}{c}
 P_{in}\\
U_{in}\\
\end{array} 
\right)
= T 
\left( 
\begin{array}{c}
 P_{out}\\
U_{out}\\
\end{array} 
\right),
\end{equation}
where
\begin{equation}
T=
\left[
\prod_{n=1}^{N_h-1}
M_n^{(h)} \, M_n^{(w)} 
\right] M_N^{(h)}
\left( 
\begin{array}{c}
 P_{out}\\
U_{out}\\
\end{array} 
\right),
\end{equation}
with
\begin{equation}
M_n^{(w)} = 
\left(
\begin{array}{cc}
 \cos (k d_n) & -i Z_c \sin(k d_n)\\
-i/Z_c \sin (k d_n) &  \cos (k d_n)\\
\end{array} 
\right) 
\end{equation}
and
\begin{equation}
M_n^{(h)} = 
\left(
\begin{array}{cc}
 1 & 0\\
1/Z_n &  1\\
\end{array} 
\right).
\end{equation}
$M_n^{(w)}$ and $M_n^{(h)}$ describe the propagation through a waveguide of length $d_n=x_{n+1}-x_n$ and through a branched scatterer with input impedance $Z_n$, respectively. $P_{in}$ ($U_{in}$) and $P_{out}$ ($U_{out}$) represent the pressure (volume velocity) at the input and output of the system, respectively. This transmission matrix method allows to determine directly the pressure field inside the structure and the transmission coefficient $t$ of the aperiodic system by \cite{Theocharis2014} 
\begin{equation}
t=\frac{2}{T_{11} + T_{12}/Z_c + T_{21} Z_c + T_{22}}.
\end{equation}

\end{document}